\documentclass[fleqn,usenatbib]{mnras}

\usepackage{newtxtext,newtxmath}

\usepackage[T1]{fontenc}

\DeclareRobustCommand{\VAN}[3]{#2}
\let\VANthebibliography\thebibliography
\def\thebibliography{\DeclareRobustCommand{\VAN}[3]{##3}\VANthebibliography}

\usepackage{booktabs}       

\usepackage{graphicx}	
\usepackage{amsmath}	
\usepackage{subcaption} 
\usepackage[export]{adjustbox} 

\title[Mitigating Bias in Deep Learning]{Mitigating Bias in Deep Learning: Training Unbiased Models on Biased Data for the Morphological Classification of Galaxies}

\author[E. Medina-Rosales, G. Cabrera-Vives and C. J. Miller]{
Esteban Medina-Rosales$^{1, 2}$,
Guillermo Cabrera-Vives$^{1, 2, 3, 4}$\thanks{E-mail: guillecabrera@inf.udec.cl}
and Christopher J. Miller$^{5}$
\\
$^{1}$Department of Computer Science, Universidad de Concepción, Edmundo Larenas 219, Concepción, Chile\\
$^{2}$Data Science Unit, Universidad de Concepción, Edmundo Larenas 310, Chile\\
$^{3}$Millennium Nucleus on Young Exoplanets and their Moons (YEMS), Chile\\
$^{4}$Millenium Institute of Astrophysics (MAS), Av. Vicuña Mackenna 4860, Macul, Santiago, Chile\\
$^{5}$Department of Astronomy and Department of Physics, University of Michigan, 1085 S. University, Ann Arbor, MI 48109, USA
}

\date{Accepted XXX. Received YYY; in original form ZZZ}

\pubyear{2023}

\begin{document}
\label{firstpage}
\pagerange{\pageref{firstpage}--\pageref{lastpage}}
\maketitle


\begin{abstract}
Galaxy morphologies and their relation with physical properties have been a relevant subject of study in the past. Most galaxy morphology catalogs have been labelled by human annotators or by machine learning models trained on human labelled data. Human generated labels have been shown to contain biases in terms of the observational properties of the data, such as image resolution. These biases are independent of the annotators, that is, are present even in catalogs labelled by experts. In this work, we demonstrate that training deep learning models on biased galaxy data produce biased models, meaning that the biases in the training data are transferred to the predictions of the new models. We also propose a method to train deep learning models that considers this inherent labelling bias, to obtain a de-biased model even when training on biased data. We show that models trained using our deep de-biasing method are capable of reducing the bias of human labelled datasets.
\end{abstract}








\section{Introduction}

Galaxies and their properties have been studied for over a century. This can be done by studying the morphology of galaxies. The seminal example of a galaxy morphological classification scheme is the Hubble sequence \citep{Hubble1926}, which distinguishes galaxies into one of four classes: elliptical, lenticular, spiral or irregular. Over the time, new classification schemes have been proposed, such as De Vaucouleurs system \citep{DeVaucouleurs1959}, which introduces new and more detailed classes. Galaxy morphologies have been shown to correlate with intrinsic properties such as gas content, brightness, color \citep{Dressler1980} and star formation \citep{Lee2013, Snyder2015}.

\begin{figure}
	\includegraphics[width=\columnwidth]{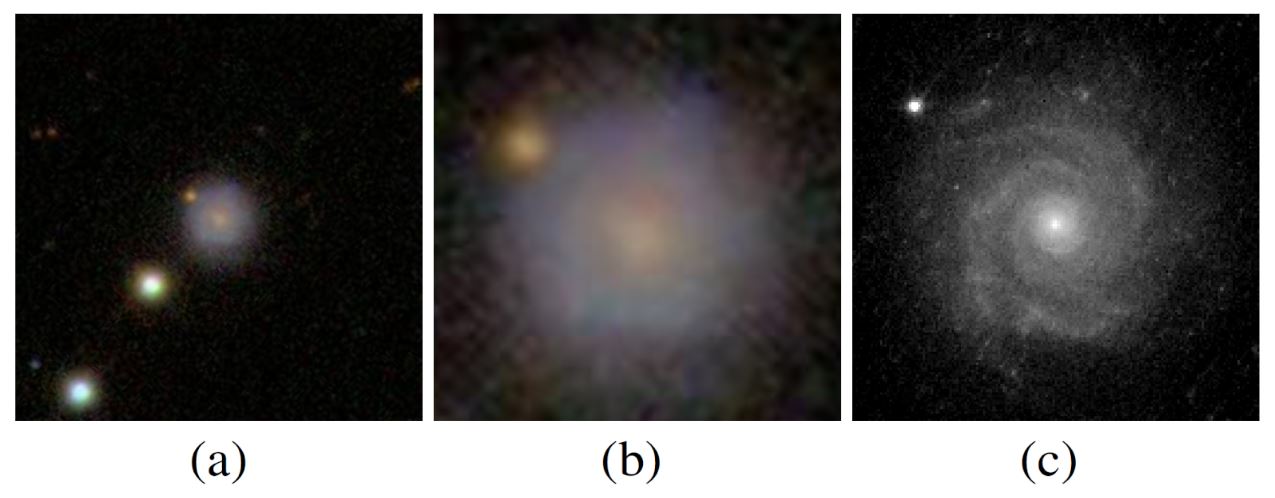}
    \caption{Example of biased classification. (a) A galaxy image as shown to the annotators, taken from the Earth's surface. (b) The same galaxy image zoomed in. (c) The same galaxy at a higher resolution taken from above the Earth's atmosphere. Notice how the features present in (c) are not clearly distinguishable in (a) and (b), even when zooming in. The majority of the annotators classified this galaxy as smooth with no signs of a disk, even though the disk features are easily distinguishable when looking at the high resolution image (c).}
    \label{fig:galaxies_context}
\end{figure}

For some time, visual classification played the dominant role in galaxy morphologies, either done by expert astronomers \citep{DeVaucouleurs1976, DeVaucouleurs1991, Bundy2005, Fukugita2007, Schawinski2007, Nair2010, Kartaltepe2015} or through crowdsourcing systems such as Galaxy Zoo \citep{Lintott2008, Bamford2009, Lintott2011, Willett2013, Simmons2017, Willett2017}. Recently, new datasets of morphologically classified galaxies obtained using diverse machine learning methods have been published, many of them using supervised learning \citep{Gauci2010, Huertas-Company2011, Dieleman2015, Huertas-Company2015, Khalifa2018, Zhu2019}, which requires labelled datasets for training. The issue is that it has been demonstrated that human-generated labels are prone to bias in terms of observable parameters. This bias causes low resolution galaxies to be skewed towards smoother types because human annotators are unable to discern the fine structures of these galaxies from the limited information provided by the low resolution images (Fig.~\ref{fig:galaxies_context}). Moreover, with the increasing popularity of deep learning models in astronomical applications and the fact that typically the only accessible labels are human-generated, the issue becomes even more critical, as these models have demonstrated a tendency to internalize human biases present in the training data.

The type of bias we address in this paper, specifically related to galaxy morphologies, has been extensively studied by the Galaxy Zoo team, \cite{Bamford2009} quantified a luminosity, size and redshift dependent bias present in Galaxy Zoo (GZ1) \citep{Lintott2008} data. They also corrected the vote fractions obtained from the crowdsourcing system by assuming that the morphological fraction does not evolve over the redshift within bins of fixed galaxy physical size and luminosity. Later, \cite{Willett2013} adapted this technique to Galaxy Zoo 2 (GZ2) data, taking in consideration that GZ2 uses a decision tree rather than a single question (as GZ1 does), meaning that all tasks but the first one depend on responses to tasks higher in the decision tree.
After that, \cite{Hart2016} presented an improved method that addresses the questions on GZ2 decision tree with multiple responses (such as number of spiral arms), showing that the method from \cite{Willett2013} does not always adjust the vote fractions correctly for this type of tasks. Their method aims to make the vote distributions consistent at different redshift rather than the mean vote fractions values as in \cite{Bamford2009} and \cite{Willett2013}. A new method to calibrate morphologies for galaxies of different luminosities and at different redshifts applied to data from the Hubble Space Telescope was introduced by \cite{Willett2017}, they used artificially redshifted images as a baseline in order to correct the galaxy morphologies.

Outside of Galaxy Zoo, \cite{Cabrera2018} introduced a metric to estimate observational bias in a dataset, also focused in galaxy morphologies. They assume that the fractions of objects of each class in an unbiased dataset should not be significantly different for labels with different resolutions within bins of intrinsic parameters. Closer to this paper, \cite{Cabrera2014} and \cite{Bootkrajang2016} address this problem through a machine learning approach, simultaneously learning a classification model, estimating the intrinsic biases in the ground truth, and providing new de-biased labels.

In this work, we propose a method to train deep learning models that takes into account this labelling bias, to obtain an unbiased model even when training with a biased dataset. Our goal is that the resulting model can be used to de-bias existing datasets and for labelling new data. We also demonstrate that deep learning models trained with biased data transfer this bias to its predictions.

This paper is organized as follows: in Section~\ref{sec:method} we introduce our de-biasing formulation and bias model. In Section~\ref{sec:results} we describe our experiments and report the results. Finally, in Section~\ref{sec:conclusions} we summarize our work and the results obtained.







\section{Methodology} \label{sec:method}

\subsection{Incorporating Bias into the Likelihood} \label{sec:likelihood}
Consider a human labelled dataset $\mathcal{D} = \{(\mathbf{x}_i, \tilde{y}_i)\}_{i=1}^N$ composed of pairs $(\mathbf{x}_i, \tilde{y}_i)$ of features $\mathbf{x}_i$ and human generated labels $\tilde{y}_i$. We assume each of these pairs to be sampled independently from a data distribution $p_{\mathrm{bias}}(\mathbf{x}, \tilde{y})$ defined over $\mathcal{X}\times\mathcal{Y}$. We also assume the existence of an unknown \emph{ground truth} label $y_i\in \mathcal{Y}$ for each biased label $\tilde{y}_i$. We consider a supervised classification task, that is, we want to approximate a function $f_\mathbf{w}:\mathcal{X}\to\mathcal{Y}$ that maps the input features to the latent ground truth labels, with parameters $\mathbf{w}$ to be fitted. For this, we consider a biasing parameter $\alpha$ (e.g. the resolution of the labelled image). In this work, we consider a classification task where $\mathcal{Y} = {1, \ldots, K}$ and use a maximum likelihood estimation approach. Considering the model parameters and the biasing parameters, the likelihood of the data can be expressed as

\begin{align}
    p(\mathcal{D}|\mathbf{w}, \{\alpha_i\}_{i=1}^N)&= \prod_{i=1}^N p(\tilde{y}_i|\mathbf{x}_i, \mathbf{w}, \alpha_i),\\
    &= \prod_{i=1}^N \sum_{y_i} p(\tilde{y}_i, y_i|\mathbf{x}_i, \mathbf{w}, \alpha_i),\\
    &= \prod_{i=1}^N \sum_{y_i} p(\tilde{y}_i|y_i, \alpha_i)p(y_i|\mathbf{x}_i, \mathbf{w}),\label{eq:marg}
\end{align}
where  the sum in equation~(\ref{eq:marg}) runs over the possible values of $y_i$. The first term of equation~(\ref{eq:marg}) models the bias in the data by assuming that the biased labels $\tilde{y}_i$ depends only on the ground truth labels $y_i$ and the biasing parameters $\alpha_i$ of each object. The second term corresponds to a probabilistic classification model used to infer the ground truth labels from features $\mathbf{x}_i$ (e.g. 2D imaging data) and the model parameters $\mathbf{w}$.
Notice that we model the bias in the data by assuming that the biased labels $\tilde{y}$ do not depend directly on the features $\mathbf{x}$. At the same time, $p(y|\mathbf{x}, \mathbf{w})$ models the dependence of $y$ with $\mathbf{x}$, making $\tilde{y}$ and $\mathbf{x}$ conditionally independent given $y$. 
In order to train a de-biased model, we minimize the negative of the log-likelihood of equation~(\ref{eq:marg}), that is, we use the de-biasing loss function
\begin{align}
    \mathcal{L} = -\log p(\mathcal{D}|\mathbf{w}, \{\alpha_i\}_{i=1}^N).\label{eq:loss}
\end{align}

\subsection{Modeling galaxy morphology labelling bias} \label{sec:bias}

In this work, we consider a binary classification task where we classify galaxies according to the first level of the Galaxy Zoo 2 classification tree \citep{Willett2013}. The first class corresponds to smooth featureless galaxies with a rounded ellipsoidal shape, also known as ellipticals. The second class corresponds to galaxies that possess a disk and present features such as spiral arms, including both spiral and lenticular galaxies. For purposes of this paper, this classes will be addressed as smooth and disk respectively. We define our labels by using Galaxy Zoo 2 weighted vote fractions corresponding to the first task of the tree, labelling galaxies as elliptical when the majority of the votes for the first task correspond to smooth, as disk when the majority of the votes correspond to features or disk and we discarded galaxies corresponding to star or artefact.


In order to model the labelling bias contained in the data, we follow \cite{Cabrera2014} and model the bias in terms of the biasing parameter $\alpha$. For this work, we define $\alpha$ as the resolution of the galaxy image measured as the angular Petrosian radius of the galaxy over the angular size of the point spread function (PSF). As mentioned before, the bias we are addressing in this work is a result of low resolution disk galaxies being mislabelled as smooth galaxies, since the distinctive features of these galaxies are not distinguishable by the annotators. We follow \cite{Cabrera2014}, and model the bias as
\begin{equation}
\label{ec:bias}
    p(\tilde{y}= \mathrm{smooth}|y= \mathrm{disk}, \alpha) = e^{-\alpha^2/(2\theta^{2})},
\end{equation}
where $\theta$ is a parameter to be fitted. Notice that $\lim_{\alpha\to 0}p(\tilde{y}= \mathrm{smooth}|y= \mathrm{disk}, \alpha)= 1$ and $\lim_{\alpha\to \infty}p(\tilde{y}= \mathrm{smooth}|y= \mathrm{disk}, \alpha)= 0$, that is, we expect low resolution disk galaxies to always be mislabelled as smooth galaxies by the annotators and, at the same time, we assume that high resolution disk galaxies are never mislabelled. On the other hand, we assume that smooth galaxies are never mislabelled as disk galaxies, this is based on the idea that a smooth featureless galaxy will preserve its smooth appearance even at low resolution, thus we define

\begin{equation}
\label{ec:bias2}
    p(\tilde{y}= \mathrm{disk}|y= \mathrm{smooth}, \alpha) = 0.
\end{equation}
From equation~(\ref{ec:bias2}) we can deduce that $p(\tilde{y}= \mathrm{smooth}|y= \mathrm{smooth}, \alpha) = 1$, as we mentioned, we do not expect smooth galaxies to be mislabelled.

\subsection{Implementing the de-biasing}
\label{sec:implement}

Now that we have shown how we can incorporate the bias model into the loss, we next consider how to implement the technique into two different classification models given data that has already been classified into smooth/disk categories by humans. The first method uses 1D catalog data for each galaxy to enable a classification. While these catalog data were measured from the 2D imaging data, the pixels themselves are not directly used in the classification. The second classification model is based on Deep Learning techniques and requires only the 2D galaxy imaging data (the pixels).

\subsubsection{DCV14 Logistic Regression}
Following \cite{Cabrera2014}, we implemented the first catalog-based model using a logistic regression as the classifier. The parameters used to make classification are the Sérsic index \citep{Sersic1968}, the ellipticity, and the half-light radius as classification features. To train this model we follow \cite{Cabrera2014} and maximize the log-likelihood introduced in Section~\ref{sec:likelihood} using the Expectation-Maximization algorithm. Our algorithm allows us to simultaneously constrain the parameters of the logistic regression as well as the parameter $\theta$ of the bias model (equation~(\ref{ec:bias})).

\subsubsection{Deep learning De-Biasing (DDB)}
\begin{figure}
\centering
\includegraphics[width=\columnwidth]{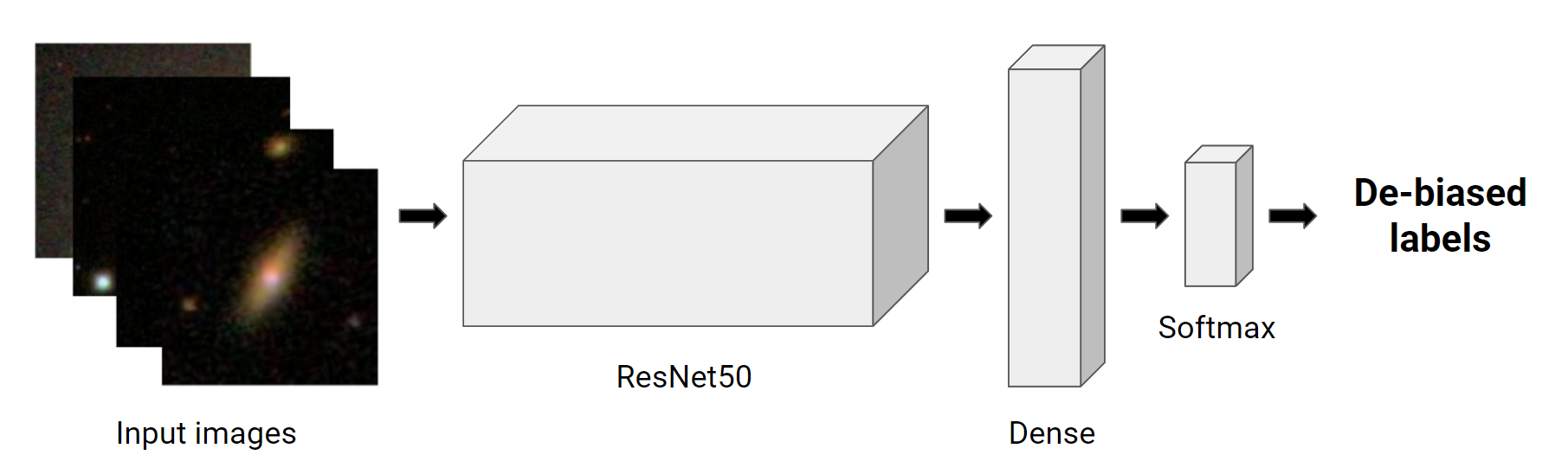}
\caption{Architecture of our deep learning de-biasing model. We used a ResNet50 with an extra dense layer and JPEG images as input. We used the negative log-likelihood described in Section~\ref{sec:likelihood} as our loss function.}
\label{fig:resnet}
\end{figure}

For our second model, we used a ResNet50 \citep{He2016} as the underlying deep learning classification model and added an additional dense layer with 1024 neurons and ReLU activation following the convolutional and pooling layers (Fig.~\ref{fig:resnet}). As input we used JPEG images and we trained our model by minimizing the negative of the log-likelihood (equation~(\ref{eq:loss})), where features $\mathbf{x}$ corresponds to the 2D galaxy imaging data and parameters $\mathbf{w}$ to the weights of the ResNet50 classification model.
When using this technique, we turn the de-biasing component on by feeding the biasing parameter $\alpha$ into our loss function. We note that $\theta$ is not included as a trainable parameter in our neural network, i.e., it is included as a constant in the loss function. In order to estimate the value of $\theta$ we use the method described in Section~\ref{sec:theta_estimation}. It is also important to note that both $\alpha$ and $\theta$ are only used during the training process, for labelling new data only the 2D galaxy imaging data is needed. 



\subsubsection{Estimating $\theta$ for DDB} \label{sec:theta_estimation}
We define the class fraction of a given class as the number of objects of said class over the total number of objects within an interval. For this work, we define the smooth class fraction $f_s = N_s/N$ and the disk class fraction $f_d = N_d/N$ in a given interval, such that $f_s + f_d = 1$ and $N_s + N_d = N$. Where $N$ corresponds to the total number of objects in the interval, $N_s$ to the total number of objects of the smooth class in the interval and $N_d$ to the total number of objects of the disk class in the interval. We also define $\tilde{f_s}$ and $\tilde{f_d}$ as their respective estimated values, that is, the values of $f_s$ and $f_d$ we obtain from the biased data.

Following \cite{Cabrera2014}, we assume that an unbiased dataset should be uniformly distributed in terms of its observable properties, i.e., $f_s$ and $f_d$ should not vary in terms of $\alpha$. Considering this, we group galaxies in bins of $\alpha$ and calculate the corresponding $\tilde{f_s}$ and $\tilde{f_d}$ for each bin.

The ground truth class fraction can be expressed as
\begin{align}
    f_s &= p(y=\mathrm{smooth}),\\
    f_d &= p(y=\mathrm{disk}),
\end{align}
and the estimated class fractions as
\begin{align}
    \tilde{f_s} &= p(\tilde{y}=\mathrm{smooth}|\alpha),\label{eq:eq5}\\
    \tilde{f_d} &= p(\tilde{y}=\mathrm{disk}|\alpha).
\end{align}
Expanding equation (\ref{eq:eq5})
\begin{align}
    \tilde{f_s}  &=  p(\tilde{y}=\mathrm{s}|\alpha),\\
    &= \sum_{y \hspace{0.2em} \in \hspace{0.2em} \{\mathrm{s, d}\}} p(\tilde{y}=\mathrm{s}, y |\alpha),\\
    &= \sum_{y \hspace{0.2em} \in \hspace{0.2em} \{\mathrm{s, d}\}} p(\tilde{y}=\mathrm{s} | y, \alpha)   p(y|\alpha),\\
    &= p(\tilde{y}=\mathrm{s}|y=\mathrm{s}, \alpha) p(y=\mathrm{s}|\alpha) + \\
    &\nonumber\hspace{1em}p(\tilde{y}=\mathrm{s}|y=\mathrm{d}, \alpha) p(y=\mathrm{d}|\alpha),
    \end{align}
where s and d correspond to smooth and disk respectively. Considering that $p(\tilde{y}=\mathrm{s}|y=\mathrm{s}, \alpha) = 1$ (smooth galaxies are always labelled as smooth) and that the ground truth label $y$ is independent of $\alpha$, we obtain
\begin{equation}
    \tilde{f_s} = f_s + f_d \cdot p(\tilde{y}=\mathrm{smooth}|y=\mathrm{disk}, \alpha).\label{eq:smooth_fraction}
\end{equation}

In order to estimate $\theta$ we minimize the square of the difference between $f_s$ and $\tilde{f_s}$ with respect to $\theta$, $f_s$ and $f_d$. This is
\begin{equation}
    \min_{\theta, f_s, f_d} \sum_i(f_s - \tilde{f}_s)^2,
\end{equation}
where the sum iterates over the bins in $\alpha$. We use equation (\ref{eq:smooth_fraction}) for $\tilde{f_s}$. Given that we do not have access to the ground truth class fractions, we initialize $f_s$ and $f_d$ using the values of $\tilde{f}_s$ and $\tilde{f}_d$ from the least biased bin in terms of $\alpha$. For the value of $\alpha$ corresponding to each bin we use the midpoint of the bin. Notice that the number of bins is a hyperparameter, for this work we grouped the galaxies in 1525 bins with the same number of galaxies in each of them.

\subsection{Quantifying the bias}
\label{sec:quant_bias}
In order to quantify the bias in a set of labels, we considered two metrics defined by \cite{Cabrera2014} and \cite{Cabrera2018}, we will refer to them as CV14 and CV18 respectively. CV14 is based on the assumption that the class fractions (same as defined in Section~\ref{sec:theta_estimation}) in a non-biased dataset should not differ in terms of the observable properties, such as image resolution. For this, they divide the dataset in bins of these observables and calculate the deviation of the class fractions as a function of them. As mentioned in Section~\ref{sec:bias}, we use $\alpha$ as our observable property (i.e., biasing parameter). For the purposes of this work, that is, a binary classification problem with only one observable property, the deviation of the class fraction for a specific class can be expressed as
\begin{align}
    \sigma_k = \sqrt{\frac{1}{N_{A}} \sum_{l=1}^{N_{A}} (\tilde{f}_{l,k} - f_k)^2},
\end{align}
where $N_{A}$ is the number of bins in $\alpha$, $\tilde{f}_{l,k}$ corresponds to the class fraction of class $k \in \mathrm{\{smooth, disk\}}$ within bin $l$, and $f_k$ to the class fraction of class $k$ from the least biased bin in terms of $\alpha$. Then, the deviation of the class fractions over the entire dataset can be quantified by
\begin{align}
    CV14 = \sqrt{\frac{1}{N_K} \sum_{k} \sigma_k^2},
\end{align}
with $N_K$ being the number of classes. For a non-biased dataset, the class fractions should be independent of $\alpha$, i.e., we expect CV14 to be close to zero.

CV18 works under the same assumption as CV14, however, it also takes into account the intrinsic properties of the data. CV18 considers that the class fractions in an unbiased dataset should not differ in terms of the observables \emph{within bins of intrinsic properties}. They first divide the dataset in multi-dimensional bins as a function of the intrinsic properties and then divide each of these intrinsic bins in bins of the observable properties. Then, they calculate the deviation of these observable class fractions as CV14 does. By doing this, they take into consideration a possible variation in the observable class fractions related to the intrinsic properties. After considering both approaches, we decided to use CV18 as the method of measuring the bias for all our experiments, as it is the more robust of the two.

\subsection{The data} 
\label{sec:data}
For all our experiments we used galaxies contained in the GZ2 catalog \citep{Willett2013}. In order to create hard labels, we focused in the first question of the GZ2 decision tree. Participants had to choose one of three possible responses, they could answer by saying that the galaxy being shown to them is "smooth", that presents "features or disk" or that corresponds to a "star or artefact" instead of a galaxy. We used majority voting according to the weighted vote fractions to define galaxies as smooth or disk, corresponding to our previously defined labels, and discarded objects that the majority of participants classified as "star or artefact". To ensure a well-balanced dataset, we conducted a random sampling of 121,984 galaxies from which 62,225 corresponded to smooth and 59,759 corresponded to disk. In order to train the models, we arranged the galaxies into different sets: a training set consisting of 97,600 galaxies, a validation set with 12,192 galaxies, and a test set containing 12,192 galaxies. As input for the models, we used the original JPEG images labelled by the annotators. Following \cite{Dieleman2015}, we cropped the original images from $424 \hspace{0.15em} \times \hspace{0.15em} 424$ to $207 \hspace{0.15em} \times \hspace{0.15em} 207$ pixels.

\section{Results}
\label{sec:results}
In the following sections, we present the quantification of whether our de-biasing technique works in practice. Since we find that it does work, we then examine more closely the de-biased data to better understand why it works. Finally, we apply a technique to explore the neural net and learn about the processes used internally by the machine learning of ResNet50.

\subsection{Bias measurement}
In Table \ref{tab:results} we present the quantified bias as described in Section \ref{sec:quant_bias}.
We first measure the bias of the labels from GZ2, both the original crowdsourced labels (GZ2B) and the de-biased labels from \cite{Hart2016} (GZ2D). In rows c and d, we then attempt to recreate the labels using ResNet50. In other words, we use the ResNet50 machine learning algorithm on training sets of the GZ2B and the GZ2D and test whether the 2D images alone are enough to recover the human-based classifications for the majority of the data. In terms of the quantified bias, the ResNet50 trained model recovers similar levels of bias to the original classifications (i.e., within the 1$\sigma$ errors determined via bootstrapping). 

In terms of accuracy and precision, ResNet50 over GZ2B reports $0.921$ and $0.921$, while for training over GZ2D the fractions are $0.896$ and $0.866$. In other words, ResNet50 alone does an excellent (but not perfect) job of recreating human labels. We find that it does better against the unmodified raw human classifications from GZ2B compared to the post-facto de-biased classifications of GZ2D. However, these differences could be due to the fact that the GZ2D is less balanced after the labels were updated by \cite{Hart2016}.

We next ran the catalog-based DCV14 de-biasing algorithm as well as the image-based ResNet50 de-biasing algorithm over the GZ2B data.  The value of the bias model parameter for GZ2B was estimated as $\theta_{\mathrm{DCV14}} = 9.18$ for DCV14, following \cite{Cabrera2014}, and as $\theta_{\mathrm{DDB}} = 11.25$ for DDB, using the method described in Section~\ref{sec:theta_estimation}.  In both cases, the de-biasing algorithms perform better than the \citet{Hart2016} GZ2D de-biasing technique. Notably, the image-only classification algorithm out-performed the catalog-based algorithm. One possible explanation for this is that the parameters we used in the DCV14 1D technique were not inclusive of all possible ones which play a role in the label determinations.

\begin{table}
    \caption{Biases for the different datasets and models implemented.}
    \label{tab:results}
    \centering
    \begin{tabular}{lc}
 
    \toprule
    Dataset / Method & Bias CV18 \\
    \midrule
    a) GZ2B \citep{Willett2013}   & $0.3696 \pm 0.0095$ \\
    b) GZ2D \citep{Hart2016}      & $0.3106 \pm 0.0108$ \\
    \midrule
    c) ResNet50 over GZ2B       & $0.3781 \pm 0.0091$\\
    d) ResNet50 over GZ2D  & $0.3258 \pm 0.0107$\\
    \midrule
    e) DCV14 over GZ2B        & $0.2994 \pm 0.0102$\\
    f) DDB (ours) over GZ2B  & $\mathbf{0.2866 \pm 0.0112}$\\
    \bottomrule
    \end{tabular}
    \label{tabla:1}
\end{table}

\subsection{Sérsic profiles analysis}
\begin{figure*}
    \begin{subfigure}{.33\textwidth}
      \centering
      \includegraphics[width=.95\textwidth]{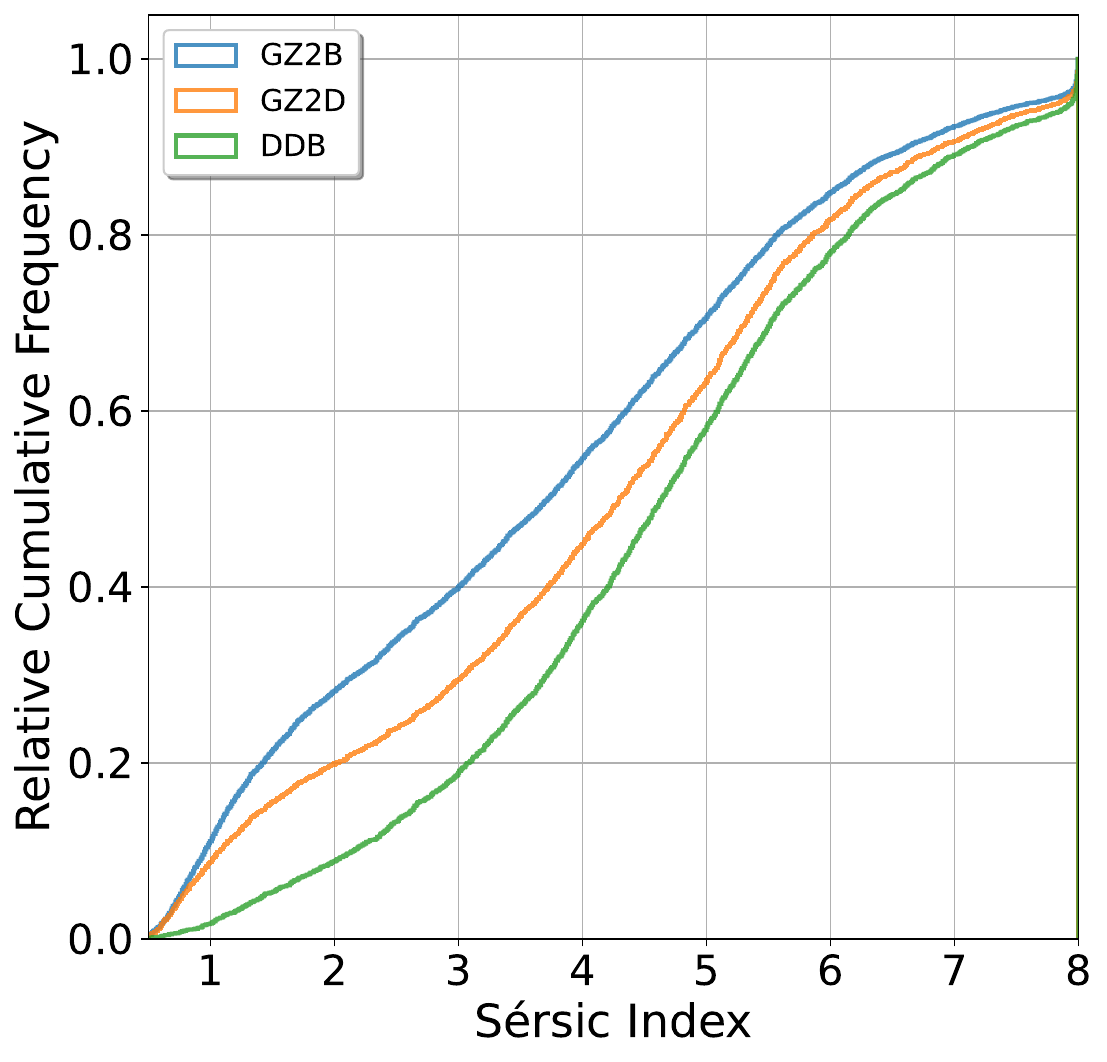}
      \captionsetup{justification=centering}
      \caption{Galaxies classified as smooth}
      \label{fig:sfig_smooth}
    \end{subfigure}%
    \begin{subfigure}{.33\textwidth}
      \centering
      \includegraphics[width=.95\textwidth]{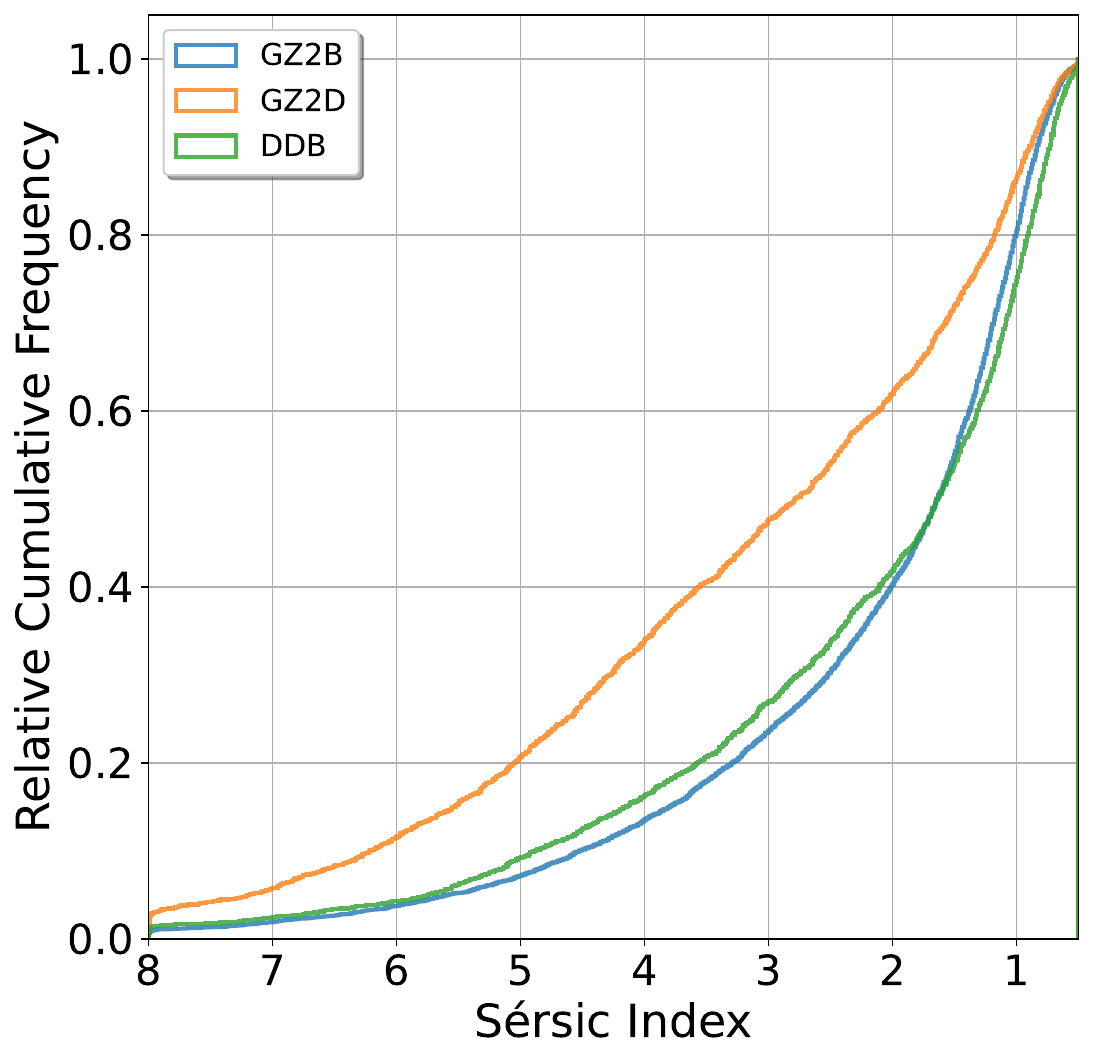}
      \captionsetup{justification=centering}
      \caption{Galaxies changed from smooth to disk}
      \label{fig:sfig_smooth_to_disk}
    \end{subfigure}
    \begin{subfigure}{.33\textwidth}
      \centering
      \includegraphics[width=.95\textwidth]{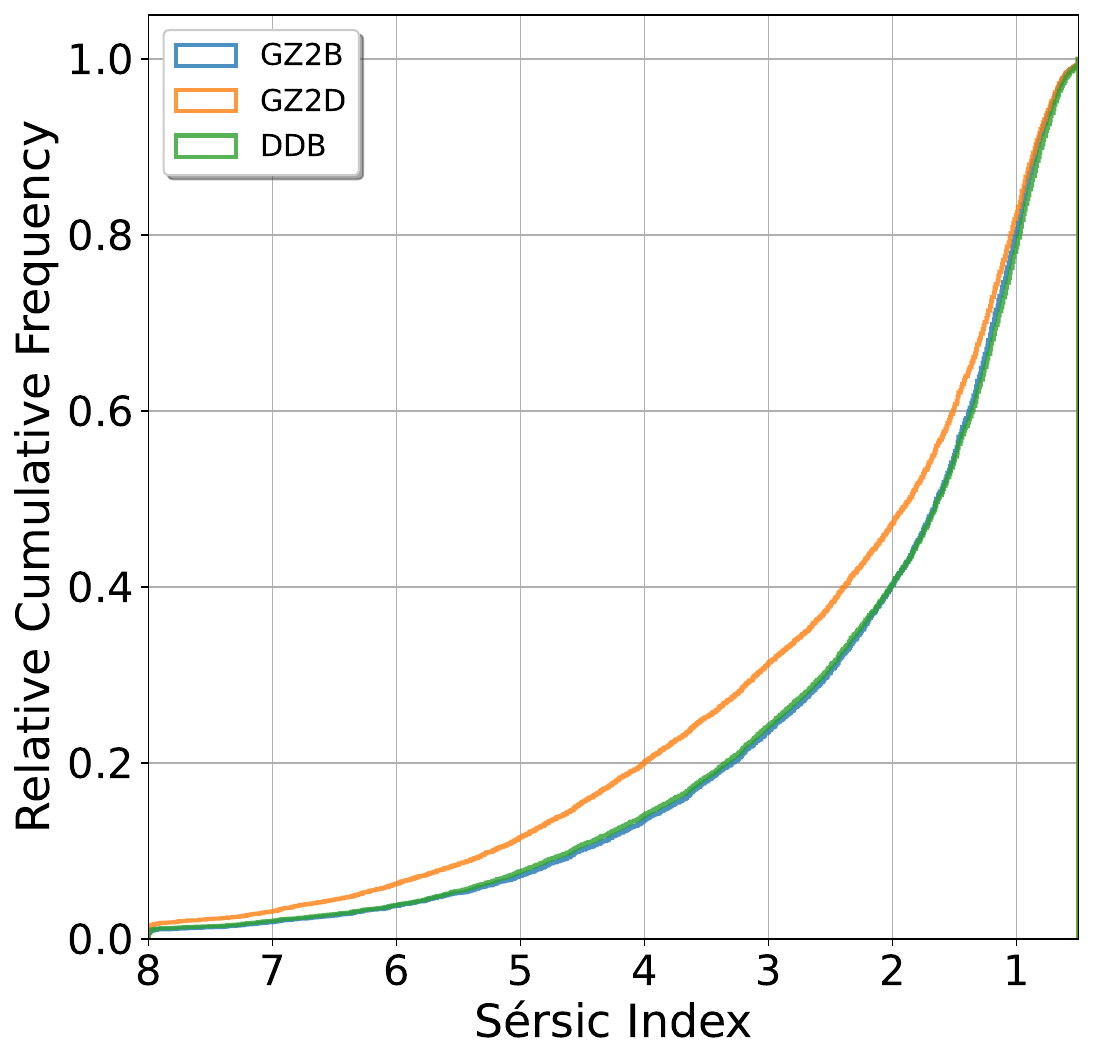}
      \captionsetup{justification=centering}
      \caption{Galaxies classified as disk}
      \label{fig:sfig_disk}
    \end{subfigure}%
\caption{Relative cumulative frequency of galaxies as classified by each model. For (b) GZ2B corresponds to the original disk galaxies, added for comparison with the other models.}
\label{fig:hist_smooth_to_disk}
\end{figure*}

To evaluate our DDB de-biasing method in terms of astrophysical parameters, we compare the cumulative frequency distribution of the Sérsic index for galaxies in our dataset (Fig.~\ref{fig:hist_smooth_to_disk}). 
A Sérsic index greater than approximately 3 is typical for smooth galaxies. In a true binary decision tree of smooth or disk galaxies, the function would resemble a step function changing from zero to one (or vice-versa) at Sérsic = 3. In reality, it should resemble an ``S''-shaped curve. In Figure~\ref{fig:sfig_smooth}, we clearly see that DDB outperforms GZ2D in terms of the Sérsic-value expectations for smooth galaxies. The biased GZ2B sample has far too many smooth galaxies (40\%) with disk-like light profiles.


On the other hand, disk galaxy light profiles are often pure exponentials with a Sérsic index of one. In Figure~\ref{fig:sfig_disk}, we see that all of the datasets do a good job of recovering the expected Sérsic index frequency. However, we notice that the GZ2D data of \citet{Hart2016} has more galaxies with higher Sérsic index values compared to the underyling biased data and also the DDB de-biased data. In
Figure~\ref{fig:sfig_smooth_to_disk}, we examine only those galaxies which switched labels from smooth to disk (by either GZ2D or DDB). Here it is unexpected to see a high relative frequency of large Sérsic index values in the GZ2D galaxies which have been labelled as disk galaxies. One explanation is that classic disk galaxies quite often have projected light profiles that resemble smooth (i.e., elliptical) galaxies. A more likely explanation is that the GZ2D re-labels are often incorrect for those galaxies which have been switched from smooth to disk by the Galaxy Zoo de-biasing algorithm.

\subsection{Visual inspection of high resolution images}
\begin{figure}
    \centering
    \includegraphics[width=\columnwidth]{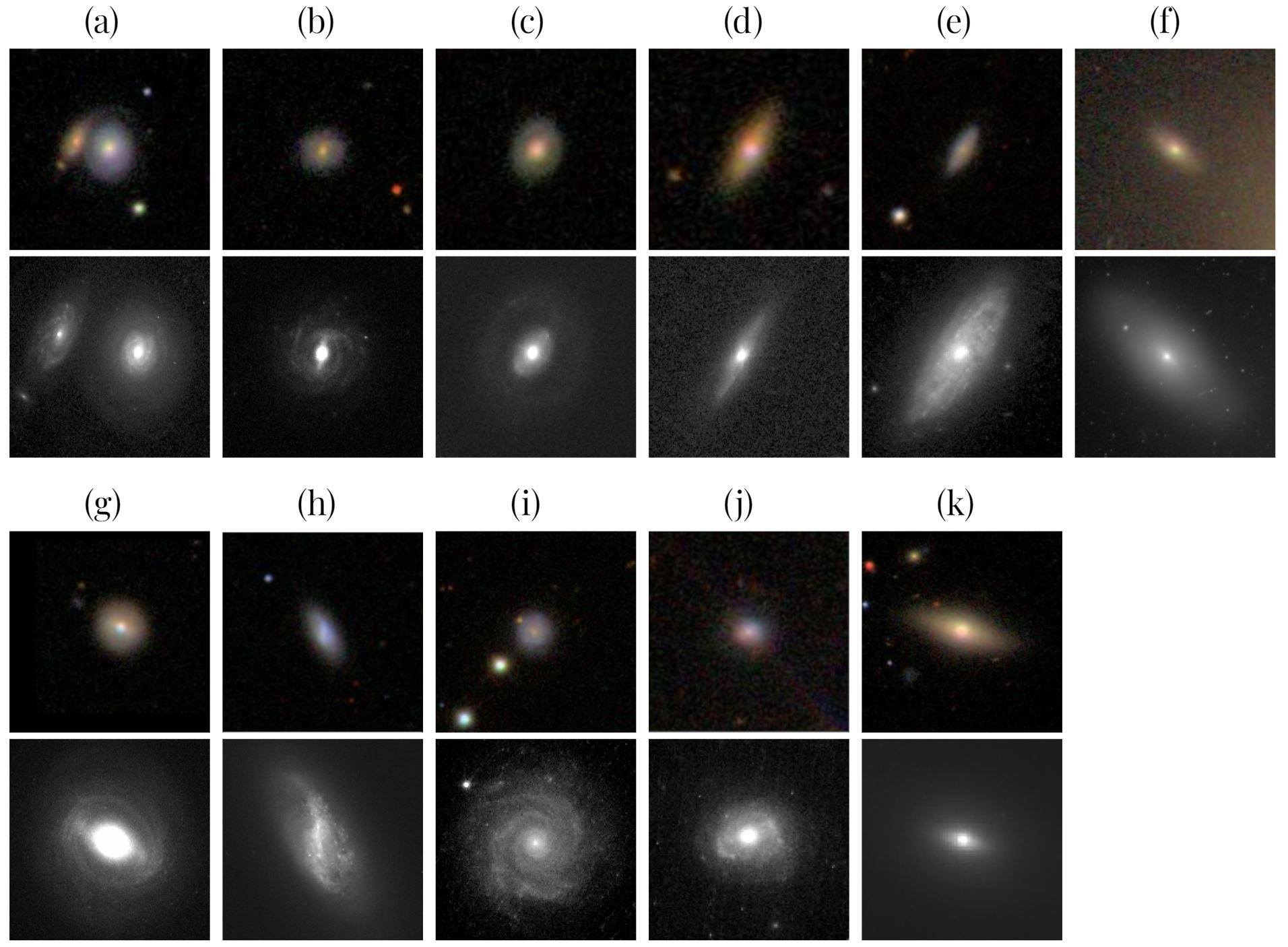}
    \caption{Galaxies that changed their labels from smooth to disk when using our method. The first row of each block corresponds to the original low resolution images as labelled by the annotators of Galaxy Zoo 2. The second row shows the higher resolution images from the HST.}
    \label{fig:galaxies}
\end{figure}

With the intention of further asses the performance of our de-biasing framework, we explored the Mikulski Archive for Space Telescopes (MAST) for high resolution images of galaxies from the Hubble Space Telescope (HST). In particular, we searched for galaxies in the test set that changed from a smooth biased label to a disk de-biased label when using DDB. We found 11 high resolution HST galaxy images that meets this criteria. Figure~\ref{fig:galaxies} shows both the original low resolution SDSS images as labelled by the annotators and their corresponding high resolution image from the HST. We notice that Figures \ref{fig:galaxies}a, \ref{fig:galaxies}b, \ref{fig:galaxies}c, \ref{fig:galaxies}e, \ref{fig:galaxies}g, \ref{fig:galaxies}h, \ref{fig:galaxies}i and \ref{fig:galaxies}j show  evidence of spiral arms, which indicates that these galaxies were correctly labelled by DDB. Figures \ref{fig:galaxies}d, \ref{fig:galaxies}f and \ref{fig:galaxies}k show lenticular features, however, it is not entirely clear that these are effectively disk galaxies. Additional astrophysical analysis is needed to validate their classification, for instance, by utilizing colors and magnitudes. These results show that DDB is able to correctly label these galaxies from the low resolution images even when the labels used for training are biased.

\subsection{GradCAM}
Aiming to understand the difference in the ResNet50 results before and after including the  de-biasing model, we employ GradCAM \citep{Selvaraju2020}, a technique for visualizing the decision making process of a convolutional neural net like ResNet50. GradCam generates a localization map that accentuates the relevant regions of the image used when predicting a specific class. 

\begin{figure}
    \centering
    \begin{subfigure}{.2989\columnwidth}
      \centering
      \includegraphics[width=.95\columnwidth]{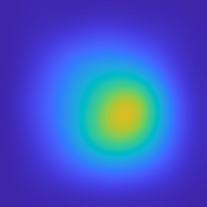}
      \captionsetup{justification=centering}
      \caption{ResNet50 over\\ GZ2B}
      \label{fig:disk_sfig1}
    \end{subfigure}%
    \begin{subfigure}{.2989\columnwidth}
      \centering
      \includegraphics[width=.95\columnwidth]{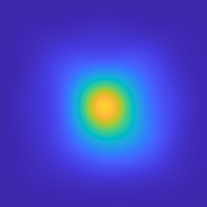}
      \captionsetup{justification=centering}
      \caption{ResNet50 over \\ GZ2D}
      \label{fig:disk_sfig2}
    \end{subfigure}%
    \begin{subfigure}{.3722\columnwidth}
      \hspace{-1.27mm}
      \centering
      \includegraphics[width=0.95\columnwidth]{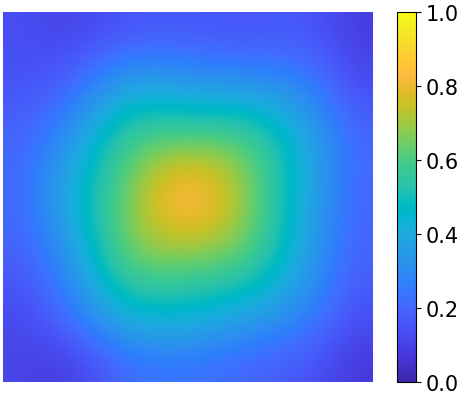}
      \captionsetup{justification=raggedright,singlelinecheck=false, format=hang, margin=0.5cm}
      \caption{DDB over \\ GZ2B}
      \label{fig:disk_sfig3}
    \end{subfigure}%
\caption{Average GradCAM localization maps for the disk class. Highlighted in yellow are the class-discriminative regions. Notice how (a) and (b) focus in the galaxy bulge while (c) considers the entire galaxy.}
\label{fig:gradcam_disk}
\end{figure}

We start by analyzing the class-discriminative regions of the models when classifying galaxies as disks. Figure~\ref{fig:gradcam_disk} shows, for each model, the average GradCAM maps of all the galaxies in the test set that the respective model predicted as a disk. Figures~\ref{fig:disk_sfig1} and \ref{fig:disk_sfig2} show that both not de-biased ResNet50 models exhibit a tendency to concentrate their attention on the centre of the image, where the bulge of the galaxy is located. On the other hand, Figure~\ref{fig:disk_sfig3} shows that DDB uses an extended area of the image, hence considering a wider fraction of the disks when discriminating.  These Figures suggest that DDB is able to detect weak disk features in the image that are often missed to the non-expert human eye.



\begin{figure}
    \centering
    \begin{subfigure}{.299\columnwidth}
      \centering
      \includegraphics[width=.95\columnwidth]{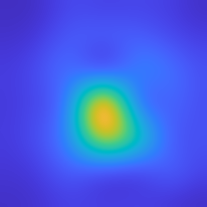}
      \captionsetup{justification=centering}
      \caption{ResNet50 over\\ GZ2B}
      \label{fig:smooth_sfig1}
    \end{subfigure}%
    \begin{subfigure}{.299\columnwidth}
      \centering
      \includegraphics[width=.95\columnwidth]{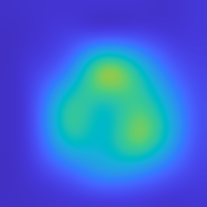}
      \captionsetup{justification=centering}
      \caption{ResNet50 over \\ GZ2D}
      \label{fig:smooth_sfig2}
    \end{subfigure}%
    \begin{subfigure}{.3717\columnwidth}
        \hspace{-1.27mm}
      \centering
      \includegraphics[width=.95\columnwidth]{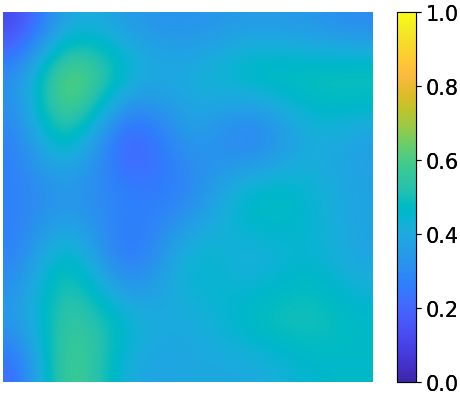}
      \captionsetup{justification=raggedright,singlelinecheck=false, format=hang, margin=0.5cm}
      \caption{DDB over \\ GZ2B}
      \label{fig:smooth_sfig3}
    \end{subfigure}%
\caption{Average GradCAM localization maps for the smooth class. Highlighted in yellow are the class-discriminative regions. Notice how (a) focuses in the bulge of the galaxy, (b) in the regions around the bulge and (c) in the the edges of the galaxy.}
\label{fig:gradcam_smooth}
\end{figure}

Figure~\ref{fig:gradcam_smooth} corresponds to the average GradCAM maps of all the galaxies classified as smooth by each model. We notice that the ResNet50 trained over GZ2B considers almost exclusively the centre of the image, which is associated with the bulge of the galaxy (Fig.~\ref{fig:smooth_sfig1}). The ResNet50 trained over GZ2D focuses mainly in the regions around the bulge (Fig.~\ref{fig:smooth_sfig2}). On the other hand, DDB concentrates its attention to a wider field within the image (Fig.~\ref{fig:smooth_sfig3}), suggesting that its decision making process is based on not identifying disk-like features in the image.
This suggests that the ResNet50 trained over GZ2D can recognize the absence of disk features near the bulge, while DDB expands its field of attention to a broader region in order to detect that no lower disk signals are present in the image when inferring that a galaxy is smooth. DDB takes into consideration the broader diversity of light profiles found among smooth galaxies, including this information in its decision making process, while both not-debiased ResNet50 models are neglecting a portion of this information by focusing mainly on the core and the regions around it.


\begin{figure}
    \centering
    \begin{subfigure}{\columnwidth}
    \centering
    \includegraphics[width=0.8\columnwidth]{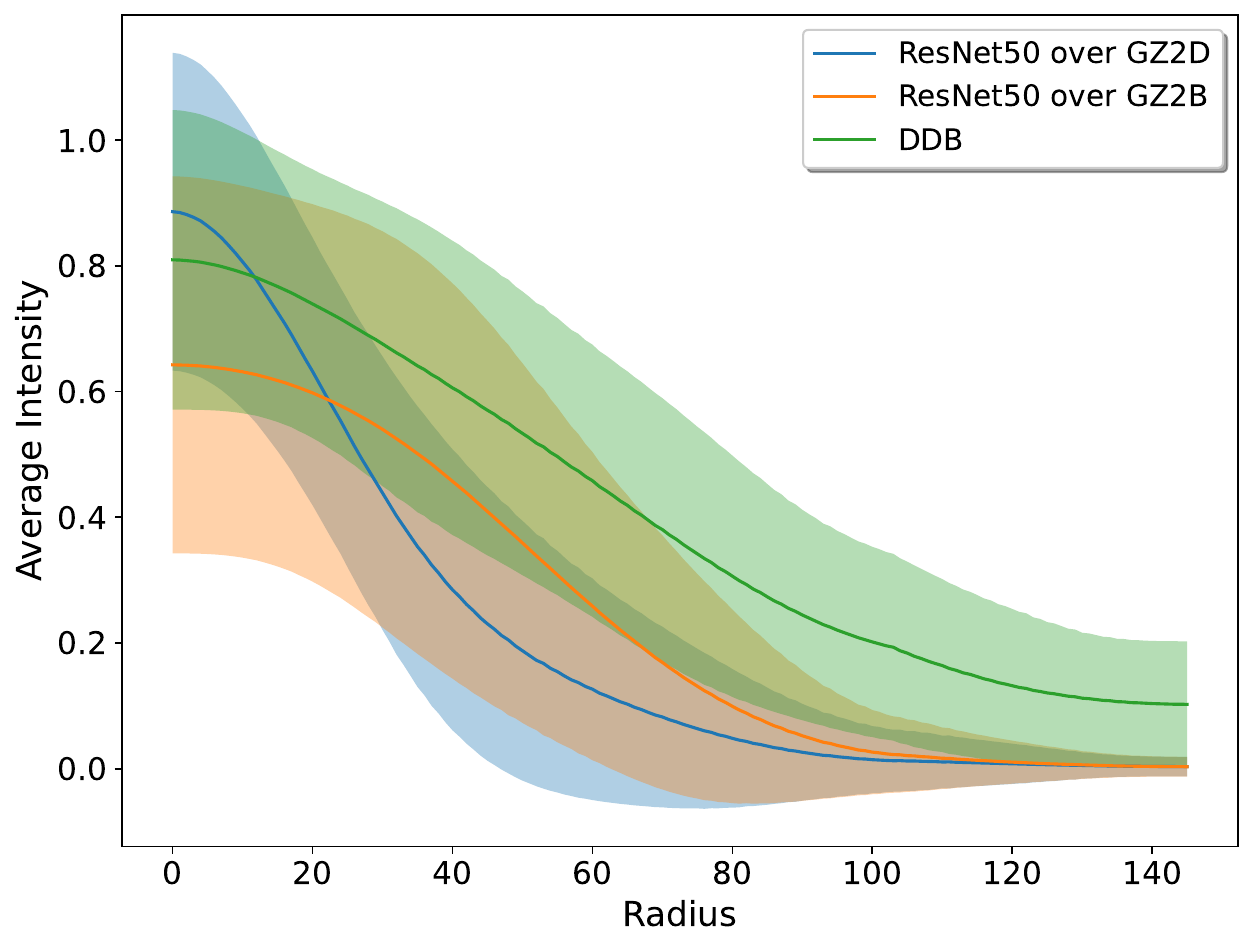}
    \caption{Disk}
    \label{fig:avg_disk_sfig}
    \end{subfigure}
    \bigskip
    \vspace{0.0005cm}
    \begin{subfigure}{\columnwidth}
    \centering
    \includegraphics[width=0.8\columnwidth]{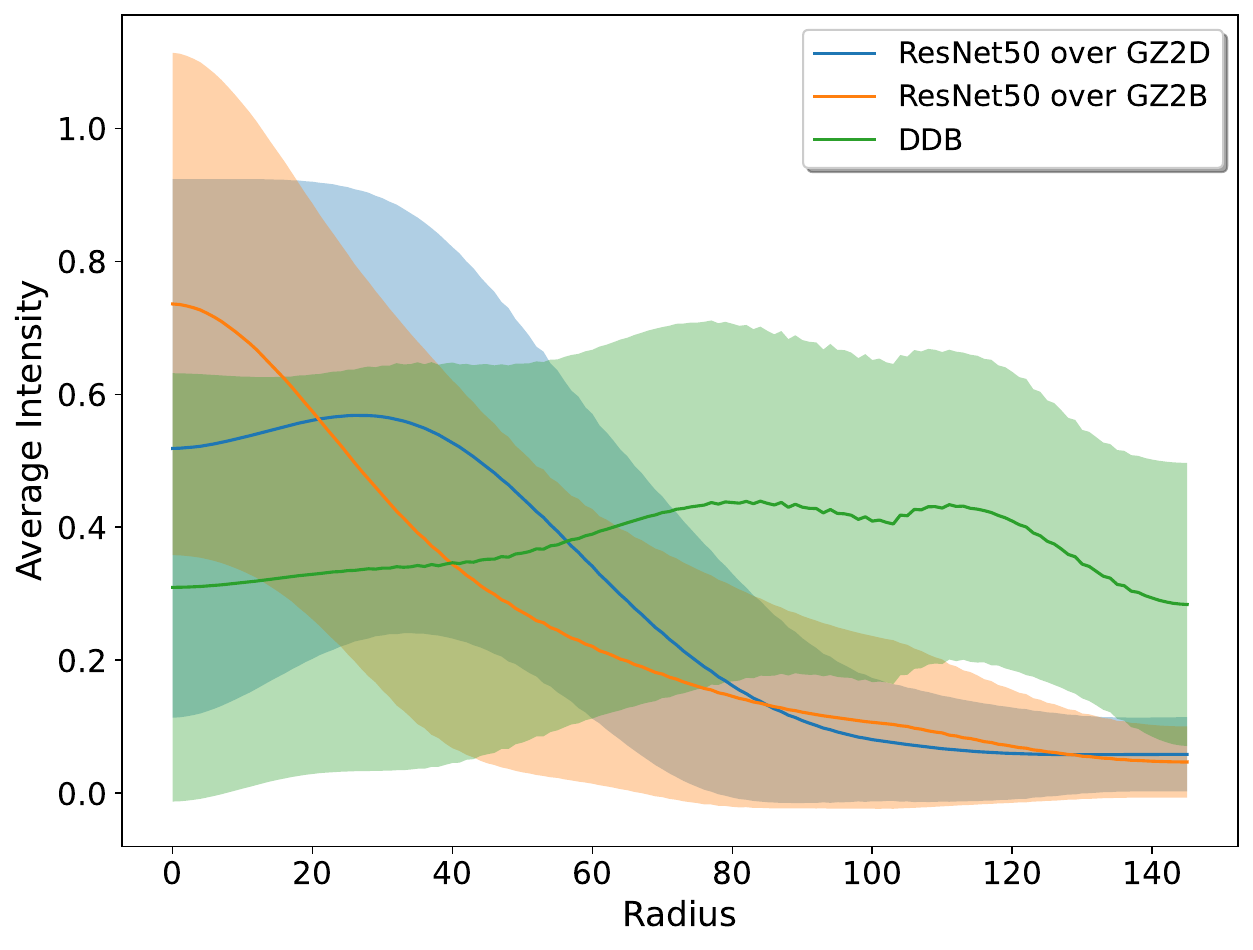}
    \caption{Smooth}
    \label{fig:avg_smooth_sfig}
    \end{subfigure}
    \vspace{0.0005cm}
    \caption{Average intensity vs. radial distance from the centre of the image. (a) Corresponds to the average GradCAM maps for the disk class (Fig.~\ref{fig:gradcam_disk}) and (b) to the average GradCAM maps for the smooth class (Fig.~\ref{fig:gradcam_smooth}).}
    \label{fig:radio_average}
\end{figure}

To provide a clearer illustration of the aforementioned findings,  Figure~\ref{fig:radio_average} shows the average intensity as a function of the radial distance from the centre of each of the GradCAM localization maps of Figures \ref{fig:gradcam_disk} and \ref{fig:gradcam_smooth}. For the GradCAM maps corresponding to the disk class (Fig.~\ref{fig:avg_disk_sfig}) we notice that DDB maintains a higher average intensity when approaching the edges, i.e., considers a broader region from the centre as relevant in its decision making process. In the case of the GradCAM maps corresponding to the smooth class (Fig.~\ref{fig:avg_smooth_sfig}) we notice that for the not de-biased ResNet50 models the average intensity decreases as the distance from the centre increases, while for DDB the average intensity remains relatively constant across the image, with its peak closer to the edges of the image.




\section{Conclusions} \label{sec:conclusions}
Data can exhibit biases in terms of observable parameters, such as resolution, which can introduce bias during the human labelling process. This inherent bias, stemming from observable properties rather than annotators, leads to systematic labelling biases in the data. In this study, we investigate the impact of training deep learning models using biased data in the context of morphological classification of galaxies. We demonstrate that training these models directly on biased data results in biased models. We introduce a method for training deep learning models that takes into account this labelling bias, to obtain unbiased models even when training with biased data. We evaluate our method by comparing the bias of the predicted labels with the bias of the labels produced by other de-biasing methods, as well as through visual inspection of high-resolution images. We also show that by using our method, the resulting model is able to learn complex astrophysical relations directly from the biased data, without relying on expert-derived parameters. Finally, we employ a visualization technique to comprehend the effect of our method on the deep learning model, comparing it with models trained without utilizing our approach. We conclude that by using our method, we can directly train a deep learning model on the biased data and obtain a model capable of both de-biasing existing datasets and labelling new data.

\section*{Acknowledgements}

The authors acknowledge support from the National Agency for Research and Development (ANID) grants: FONDECYT iniciación 11191130 (EMR, GCV); FONDECYT regular 1231877 (GCV); Millennium Science Initiative Program – NCN2021\_080 (GCV) and ICN12\_009 (GCV).

\section*{Data Availability}
The galaxy images used for training the deep learning models are from \cite{Willett2013} and are publicly available. All galaxy parameters used are from the Sloan Digital Sky Survey (SDSS) Data Release 7 (DR7) and can be obtained from the SDSS database. Data related to the Sérsic profiles was obtained from \cite{Simard2011} and is freely accessible. The images from the Hubble Space Telescope (HST) are publicly accessible from the Mikulski Archive for Space Telescopes (MAST) website: https://mast.stsci.edu/portal/Mashup/Clients/Mast/Portal.html.
 



\bibliographystyle{mnras}
\bibliography{references} 








\bsp	
\label{lastpage}
\end{document}